\documentclass[12pt]{article}
\usepackage{epsfig, graphics}

\usepackage{amsfonts}
\usepackage[latin1]{inputenc}
\newcommand{\nin}{\noindent}

\def\bkR{{\rm I\kern-.17em R}}

\def \1n{1\hskip -3pt \mbox{N}}

\newfont{\bbf}{cmbx12 scaled 1435}

\newcommand{\be}{\begin{equation}}
\newcommand{\ee}{\end{equation}}

\renewcommand{\theequation}{\thesection.\arabic{equation}}

\newcommand{\ba}{\begin{eqnarray}}
\newcommand{\ea}{\end{eqnarray}}
\newcommand{\ban}{\begin{eqnarray*}}
\newcommand{\ean}{\end{eqnarray*}}

\newcommand{\bt}{\begin{tabular}}
\newcommand{\et}{\end{tabular}}
\newcommand{\btb}{\begin{tabbing}}
\newcommand{\etb}{\end{tabbing}}
\newcommand{\bfr}{\begin{flushright}}
\newcommand{\efr}{\end{flushright}}

\newcommand{\bge}{\begin{enumerate}}
\newcommand{\ene}{\end{enumerate}}
\newcommand{\bc}{\begin{center}}
\newcommand{\ec}{\end{center}}

\def \1n{1\hskip -3pt \mbox{N}}

\begin{document}
\setlength{\baselineskip}{.26in}
\thispagestyle{empty}
\renewcommand{\thefootnote}{\fnsymbol{footnote}}
\vspace*{0cm}
\begin{center}

\setlength{\baselineskip}{.32in}
{\bbf Analysis of Virus Propagation: A Transition Model Representation of Stochastic Epidemiological Models}

\setlength{\baselineskip}{.26in}
\vspace{0.5in}
C. Gourieroux \footnote[1]{University of Toronto,  Toulouse School of Economics and CREST,{\it e-mail}:
{\tt gouriero@ensae.fr}},
 J. Jasiak  \footnote[3]{York University, Canada, {\it e-mail}:{\tt jasiakj@yorku.ca}\\
The authors gratefully acknowledge financial support of the Chair ACPR: Regulation et Systemic Risks, the ANR (Agence Nationale de la Recherche) and the Natural Sciences and Engineering Council of Canada (NSERC). We thank A. Monfort for helpful comments.}
\vspace{0.1in}
\vspace{0.5in}
  
first version: March 31, 2020\\
revised: \today\\

\vspace{0.3in}
\begin{minipage}[t]{12cm}
\small
The growing literature on the propagation of COVID-19 relies on various dynamic SIR-type models (Susceptible-Infected-Recovered) which yield model-dependent results. For transparency and ease of comparing the results, we introduce a common representation of the SIR-type stochastic epidemiological models. This representation is a discrete time transition model, which allows us to classify the epidemiological models with respect to the number of states (compartments) and their interpretation. Additionally, the transition model eliminates several limitations of the deterministic continuous time epidemiological models which are pointed out in the paper. We also show that all SIR-type models have a nonlinear (pseudo) state space representation and are easily estimable from an extended Kalman filter.\\

\end{minipage}

\end{center}
\renewcommand{\thefootnote}{\arabic{footnote}}
\newpage

\newpage

\setcounter{page}{1}

\section{Introduction}

For almost 100 years following the introductory article of Kermack, McKendrick (1927), the SIR (Susceptible-Infected-Recovered) model has remained the primary tool of analysis for epidemiological studies and has inspired a considerable number of extensions [see e.g. Hethcote (2000), Brauer,Castillo-Chavez (2001),Vynnicky et al. (2010), for general review]. In recent applications to the Coronavirus propagation however, these SIR-type models tend to 
produce results and forecasts that lack robustness and show variation across models.  

This paper introduces a common representation of SIR-type stochastic epidemiological models to facilitate the comparison between models and their outcomes. This representation is a discrete time transition model, which is used to define a typology of epidemiological models with respect to the number of states (compartments) and their interpretation. 

The discrete time transition model is characterized by a transition matrix, which determines the probabilities of transitions between the states distinguished in an epidemiological model. As such, it can easily accommodate individual and aggregate count data sampled at various frequencies. On the contrary, a discretization bias arises when a continuous time 
deterministic differential system is adapted to data sampled at fixed intervals.
In particular, the discretization bias affects the estimated collective immunity ratio, which causes its reliability and even its existence to be questionable.  Moreover, a time discretized SIR-type model is shown to provide different results in an application to data sampled at different frequencies, such as the daily or weekly frequencies, due to its inconsistency with respect to the time unit. The same limitation concerns the reproduction number, which is a commonly used epidemiological parameter. We also comment on the constrained specifications of some SIR-type models, with the potentially complex dynamics of infection probabilities being determined from very few parameters. The transition model allows us to avoid these and other limitations revealed in the commonly used epidemiological models and discussed in the paper.

When aggregate counts are observed, all SIR-type models are shown to have a (Gaussian) (pseudo) nonlinear state space representation, which is convenient for statistical inference. We show that a quasi-maximum likelihood (QML) estimation method can be applied to the pseudo state space representation and easily implemented with an extended, or unscented Kalman filter used for approximating the unobserved state probabilities.

The paper is organized as follows.
The stochastic transition model is introduced in Section 2.  First, we define the stochastic framework of individual histories, which is next transformed into a deterministic dynamic model, for the cross-sectional count aggregates over a large number of individuals.  Section 3 examines the features of the 2-state SI and 3-state SIR models. We perform the sensitivity analysis to see how the peak of new infections and the time-to-peak depend on the propagation parameters. In Section 4, the (pseudo) state space representation of an epidemiological is derived for statistical inference. In Section 5, we discuss the case when the propagation parameters display either a deterministic, or stochastic variation over time. We show that these extensions of the model can also be examined in a (pseudo) state space framework. Section 6 concludes.
The typology of SIR type models with 2, 3 and 4 states is given in Appendix 1. Proofs are gathered in Appendix 2.

\section{Contagion Modelling}

This section introduces a general specification that encompasses the main epidemiological SIR-type models. It is a discrete time stochastic transition model that allows for modelling of individual histories during an epidemic without the limitations revealed in the deterministic epidemiological models.
In particular, we point that the time discretized version of a continuous time SIR-type model depends on the time unit and needs to be re-adjusted for the sampling frequency of the data. Moreover, the reproduction number computed from the time discretized model depends on the time unit as well and takes a different value when computed from daily or weekly data. We also show a difficulty with inference based on a continuous time model of frequencies.

\subsection{The Stochastic  Transition Model}

We consider a large panel of individual histories $Y_{i,t},i=1,..,N, t=0,..,T$, where the variable $Y$ is qualitative multinomial with $J$ alternatives denoted by $j=1,...,J$. These alternatives are the states of infection, recovery or death, depending on the model specification (or compartments in the epidemiological terminology). The discrete time $t$ is assumed to be measured in days, as daily data are often used in epidemiological studies.

\medskip
\nin {\bf Assumption A1:} The individual histories are such that:

i) The variables $Y_{i,t},i=1,..,N$ at time $t$ fixed have the same marginal (i.e. cross-sectional) distributions. This common distribution depends on time $t$ and is discrete. It is determined by the vector $p(t)$ of size $J$ with components:

$$ p_j(t) = P(Y_{i,t}=j),\; j=1,..,J.$$

\nin These components are non-negative and sum up to 1 .

ii) The processes ($Y_{i,t},t=1,.,T),i=1,..,N $ are independent, heterogeneous  Markov processes of order 1, with common transition probabilities. The transitions from date $t-1$ to date $t$ are characterized by the $J \times J$ transition matrix $\Pi[p(t-1)]$. This matrix has nonnegative elements and each of its rows sums up to 1.

\medskip
The vector $p(t)$ represents the cross-sectional (marginal) probabilities of states. In practice, the cross-sectional probability $p(t)$ is close to the cross-sectional frequency $f(t)$, computed from the values of $Y_{i,t}$. Then, transition matrix $\Pi[p(t-1)]$ is close to $\Pi[f(t-1)]$. However the transitions between states have to be defined with respect to $p(t-1)$ to remain independent of the population size.  

\medskip
\nin {\bf Assumption A2:} i) The epidemic starts at time 0.
ii) At time 0 all individuals are in state $j=1$, which is interpreted as the susceptible compartment.

\medskip

Under Assumptions A1 and A2, the individual histories $Y_{i,t}$ are independent and identically distributed. Therefore, the individuals are exchangeable, i.e. have similar risk factors (homogenous population). 

The initial condition also implies nonstationary evolutions of the processes of individual histories over time. This nonstationarity and the time dependence of the transition matrix through $p(t-1)$ only are the distinct characteristics of a SIR-type model. There is one exception, however. When the SIR model is a homogenous Markov process, the transition matrix $\Pi$ is independent of $p(t-1)$.
In general however, the transition matrix $\Pi$  is time dependent. Then, the type of epidemiological model is determined by the number of states, their interpretations, and the structure of the transition matrix. More specifically, the elements of the transition matrix can be either zeros, constants, or functions of marginal probabilities $p(t-1)$ in the presence of time dependence. The models differ with respect to the form of those functions and of the components of $p(t-1)$, which are their arguments. 

The examples of commonly used SIR-type specifications are described in Appendix 1. Although most of the SIR-type models are heterogeneous Markov models, the homogeneous Markov model mentioned above can be used for either the local analysis (see Section 2.2), or for deriving the lower and upper bounds on the trajectories of marginal probabilities $p(t)$. Those bounds are mainly determined by the maximum (resp. minimum) of  moduli of all eigenvalues of $\Pi[p(t-1)]$, called the Lyapunov exponents, over time $t$.

\subsection{The Deterministic Model}

Assumptions A1 and A2 defining the stochastic dynamics of $Y_{i,t}$ lead to a deterministic nonlinear recursive model for the dynamics of marginal probabilities $p(t)$. This deterministic representation is obtained by applying the Bayes formula and is given by:

\begin{equation}
p(t)= \Pi \, [p(t-1)]' p(t-1), \; t=1,..,T,     
\end{equation}

\nin with initial condition: $p(0)= (1,0,..,0)'$. As shown in Section 4, system (2.1) can be used as a system of estimating equations for statistical inference.

System (2.1) can be rewritten to define the dynamics of changes in marginal probabilities:

\begin{equation}
\Delta p(t) = p(t)-p(t-1) = \{ \Pi \,[p(t-1)]- Id \} p(t-1),
\end{equation}

\nin where $Id$ denotes the identity matrix. The equation (2.2) highlights the role of the generator: $\Pi \,[p(t-1)]- Id$, in determining the changes in marginal probabilities $\Delta p(t)$.

\medskip
Remark 1: Discretization bias.

A major part of literature concerns epidemiological models written as deterministic differential systems in continuous time. A continuous time analogue of the deterministic model (2.2) is:

\begin{equation}
dp(t)/dt = \{ \Pi [p(t)] - Id \} p(t) .
\end{equation}

\nin In general, the system of equations (2.2) is not the exact time discretized version of the continuous time system
(2.3) [see Appendix 2 for the rational recursive system]. Thus, due to the nonlinearities in the dynamics, a chaos effect can arise and induce considerably different evolutions of $p(t)$ defined from (2.2) and (2.3), especially over the medium run.

To highlight the differences between the discrete and continuous time modelling, let us consider the 2-state SIS model with a linear force of infection, i.e. a linear function $\pi$ (see Remark 1). The probability of being infected $p_2(t)$ satisfies
the following recursive equation \footnote{Coefficients $\beta, \gamma$ are assumed constrained to ensure that $p_2(t)$ takes values between 0 and 1, for any $p_2(t-1)$ in [0,1].}:

$$ p_2(t)=[\beta p_2(t-1)] [1-p_2(t-1)] + (1- \gamma) p_2(t-1),$$

\nin in discrete time and

$$ d p_2(t)/dt= (\beta- \gamma) p_2(t) - \gamma p_2(t)^2, $$

\nin in continuous time.
Even though both equations look similar and contain the same parameter symbols, the following differences can be pointed out:

\medskip
\nin i) The discrete time version of SIS is not time consistent, as $p_2(t)$ at lag 2 derived by recursive substitution  is a quartic function of $p_2(t-2)$. Hence, this specification needs to be modified whenever the unit of time separating the observations changes. In practice, this means that a specification valid for daily data is not valid for  weekly data. On the contrary, the continuous time version of SIS is time consistent.

\medskip
\nin ii) The parameters $\beta$ and $\gamma$  in both the discrete and continuous time SIS models given above depend on the selected time unit too. In the continuous time model however, $\beta$ and $\gamma$ are multiplied by the same factor when the time unit is changed. Then, the so-called reproduction number $R_0= \beta/\gamma$ is invariant with respect to the time unit. As the discrete time model is not time consistent, $R_0$ computed directly from the time discretized model is not invariant with respect to the time unit either. Hence, different values of $R_0$ are obtained from daily and weekly data. Any result obtained from the time discretized  differential equation (called the Euler discretization)  and interpreted in the continuous time framework needs to be interpreted with caution. This finding calls into question the reliability of the reproduction number $R_0$ and some of its transforms, such as the asymptotic value $p_2(\infty)$, which are important components of epidemiological studies.

\medskip
 Remark 2: Non-differentiability of a continuous time frequency model
 
It is common to write the epidemiological model as a differential system  of frequencies $f(t)$,  instead of marginal probabilities $p(t)$, as:

\begin{equation}
df(t)/dt= [\Pi [f(t)] -Id] f(t).     
\end{equation}

\nin This differential system is not compatible with the set of admissible values of vectors $f(t)$ \footnote{It is also the case when a stochastic feature is introduced by replacing the deterministic differential equation by a stochastic one, such as a multivariate Jacobi process to account for the positivity and unit mass restrictions on the components of $f(t)$ [see,  Admani et al. (2018), Jiang et al. (2011), El Koufi et al. (2019) for examples of stochastic differential epidemic models and Gourieroux,Jasiak (2006) for the multivariate Jacobi process].}. The components of  $f(t)$ are not continuously valued functions, as they take on values equal to the multiples of $1/N$, which implies the non-differentiability of function $f(t)$. Therefore, an epidemiological model of this type cannot provide accurate results. Moreover, as the model is deterministic, it cannot take into account the ex-ante uncertainty about the vectors $f(t)$, which are random.

\subsection{Local Expansions}

Let us now examine the dynamics of marginal probabilities of states $p(t)$.
The analysis of their evolution during an epidemic can be simplified if we focus on either the beginning, or the end of the epidemic and consider local expansions.

\medskip
\nin {\bf 2.3.1 Beginning of the epidemic}

At time $t=0$, the initial value is: $p(0)=(1,0,..,0)'$. Below, we consider expansions of orders 1 and  2 of the recursive system (2.1) in a neighbourhood of $p(0)$.

\medskip
\nin {\bf i) First-Order Expansion}:

The first-order expansion  is:

\begin{equation}
p(t)= \Pi[ p(0)]' p(t-1),  
\end{equation}

\nin which corresponds to a homogeneous Markov model with transition matrix $\Pi [p(0)]$. It can be solved analytically as:

\begin{equation}
p(t)= \Pi \,[p(0)]^{'t} p(0), 
\end{equation}

\nin We find that locally the components of marginal probabilities $p(t)$ are combinations of exponential functions (and also of sine functions, cosine functions, which can possibly be multiplied by polynomials, if some eigenvalues of $\Pi \,[p(0)]$ are complex and/or multiple). Their dynamics are constrained by the specific form of matrix $\Pi[p(0)]$, which is a transition matrix. More specifically, all components of $p(t)$ have to take values between 0 and 1. In order to satisfy this restriction, locally, the marginal probability $p_1(t)$ is exponentially decreasing over time, whereas marginal probabilities of other states $p_j(t),\; j=2,..,J$ are exponentially increasing over time.

\medskip
\nin {\bf ii) Second-Order Expansion}

The second order expansion leads to the dynamic system:

\begin{equation}
p(t)= \{ \Pi \,[p(0)] + \sum_{j=1}^J d \Pi \,[p(0)]/d p_j \, p_j(t-1) \} ' p(t-1),    
\end{equation}

\nin which is a Riccati quadratic recursive system.

\nin {\bf 2.3.2 End of the epidemic}

In general, the epidemiological models include some absorbing states, such as the states of deceased, or recovered (see the examples in Section 3 and Appendix 1). In this case, the sequence of marginal probabilities $p(t)$ has a limit when $t \rightarrow \infty: p(\infty)$, say. If there is only one absorbing state $J $, say, we get
$p(\infty)= (0,0,..,1)'$. Then, the first- and second-order expansions  can be performed in a neighbourhood of $p(\infty)$, yielding dynamic approximations systems analogous to systems (2.5) and (2.7).

The  early phases of the epidemic, which are characterized  by expansions given in Section 2.3.1. are illustrated in
Section 4. Expansions derived in Section 2.3.2 can be used
for other research objectives, such as determining the level of collective immunity.

\setcounter{equation}{0}\def\theequation{3.\arabic{equation}}

\section{Examples}

Let us now study the dynamic properties of two commonly used epidemiological models, which are the SI and SIR models (see Appendix 1) in the framework of a discrete time transition model. We derive the dynamic equations of marginal probabilities and describe their behavior at the beginning and the end of an epidemic.

\subsection{SI Model}

\nin {\bf 3.1.1 The deterministic model}
\medskip

\nin The transition model representation of the deterministic SI model involves the following $2 \times 2$ transition matrix:

row 1, S: $1-\pi(p_2)$; $\pi(p_2)$.

row 2, I: 0, 1.

\nin The state I of infected, still infectious and immunized is the absorbing state. Function $\pi$ is the contagion function (called the {\bf force of infection}) that satisfies the following assumption:

\medskip
\nin {\bf Assumption A3:}  $\pi$ is a non-decreasing function of $p_2$, which takes  values between 0 and 1.

\medskip
The value $\pi(0)$ can be interpreted as an exogenous component of the contagion. In an open economy, it can be due to the effect of tourism, international trade and migration. In a closed economy, such as the world in its entirety, it can be set equal to zero. There is a {\bf strict (endogeneous) contagion} effect if function $\pi$ is strictly increasing.

\medskip
\nin {\bf Example 1:}  A common specification of the force of infection $\pi$ is the linear function: $\pi(p_2)=b \, p_2$, where parameter $b$ takes values between 0 and 1, or a logistic function of $p_2$:

$$\pi(p_2)= \exp(a+b p_2)/[ 1+ exp(a+b p_2)],$$

\nin where coefficient $b$ is non-negative. In the logistic force of infection the exogenous infection rate is measured by: $\exp a/[1+ \exp a]$ and the strict endogenous contagion effect by parameter $b$. Other functional forms have also been considered in the growth literature and obtained, for instance, by replacing  $p_2$ by a power of $p_2$ in the expressions given above [see e.g. Richards(1959), Kuhi et al. (2003), Table 1, Brandenburg (2019), Wu et al. (2020), Harvey, Kattuman (2020)].

The form of the transition matrix given above leads to the following nonlinear recursive equation of order 1 for the marginal probability of being infected:

\begin{equation}
p_2(t)= \pi(p_2(t-1))[1-p_2(t-1)] + p_2(t-1).         
\end{equation}

\medskip
\nin {\bf Proposition 1:}  Under Assumption A3, $p_2(t)$ is a non-decreasing function of time with exponential lower and upper bounds:

$1-[1-\pi(0)]^t \leq p_2(t) \leq 1-[1-\pi(1)]^t$.

\nin It tends to 1 when $t \rightarrow \infty$.
\medskip

\nin Proof: 

\nin i) It is non-decreasing, as $p_2(t) - p_2(t-1)$ is non-negative.

\nin ii) The bounds are obtained by observing that $p_2(t)$ is an increasing function of $\pi$.

\nin iii) Since $p_2(t)$ is non-decreasing and bounded by 1, it converges to a value $p_2(\infty)$. This limit is equal to 1, by considering (3.1) at $t=\infty$.

\nin QED

\medskip
In particular, if there is no contagion effect i.e. if $\pi(p_2)$ is constant and equal to $\pi$, then the marginal probability of being infected is: $p_2(t) =1-[1-\pi]^t$.

\medskip
\nin {\bf 3.1.2 Expansions}
\medskip

It is interesting to consider the expansions of the dynamics of the probability of being infected in the SI model at the beginning of the  contagion,i.e. when $p_2$ is close to zero, or at the end of the contagion, i.e. when $p_2$ is close to 1.

\medskip
\nin i) Beginning of the contagion

A second-order expansion leads to:

\begin{equation}
p_2(t)-p_2(t-1) \propto [\pi(0)+d \pi(0)/dp \, p_2(t-1)] [1-p_2(t-1)].  
\end{equation}

\medskip
\nin ii) End of the contagion

The second-order expansion in a neighbourhood of $p_2=1$ yields:

\begin{equation}
p_2(t)-1= [\pi(1)+d \pi(1)/dp \, [p_2(t-1)-1]] [1-p_2(t-1)] + p_2(t-1)-1.    
\end{equation}

Both approximations lead to discrete time logistic recursive equations for the probability of being infected $p_2(t)$ and the probability of not being infected  $1-p_2(t)$, respectively.

\medskip
\nin {\bf 3.1.3 Continuous time analogue}
\medskip

\nin The continuous time analogue of the recursive equation (3.2) is:

\begin{equation}
d p_2(t)/dt= (\alpha + \beta p_2(t)) (1- p_2(t)),           
\end{equation}

\nin where $\alpha= \pi(0), \beta= d \pi(0)/ dp$ are both nonnegative.

\nin Equation (3.4) can be solved analytically.

\medskip
\nin {\bf Proposition 2:}  i) Assuming that the beginning of the epidemics is at time $t=0$, the solution of equation (3.4) is:

$$p_2(t)= [\alpha \exp[(\alpha + \beta)t]- \alpha]/[\alpha \exp[( \alpha+ \beta) t] + \beta].$$

ii) If $\beta > \alpha$, the solution is such that the derivative $d p_2(t)/dt$ attains the  maximum when $p_2(t)= (\beta- \alpha)/(2 \beta)$. The time-to- inflection is reached at $t^*= \log (\beta/\alpha)/(\alpha+\beta)$.
\medskip

\nin Proof: see Appendix 2.

\medskip
It follows that the probability of being infected $p_2(t)$ is a logistic function of time. Moreover, if the strict contagion effect is large as compared to the exogenous component of the contagion, there is a peak in the changes of ratios of infected individuals over time. The size and timing of the peak depend on the propagation parameters. However, the SI model has only two parameters, which is insufficient to independently determine the peak, the time to peak and other characteristics such as the flatness of the curve at the peak (the so-called "plateau" effect) as well as the asymmetry of the curve with respect to the peak.

These above outcomes of the SI model (3.1) are based on a local expansion of the initial nonlinear recursive equation and are therefore valid at the beginning of an epidemic only. The length of the time episode over which such an expansion is valid depends on  function $\pi$, and also on the values of parameters $a, b$ in the parametric SI model in Example 1.

\medskip
\nin {\bf 3.1.4 Sensitivity Analysis}

\medskip

Let us consider below the parametric SI model in Example 1 and illustrate graphically its dynamics. At time 0, we  fix the probability of being infected $p_2(0)=0$ and set the parameters $\alpha, \beta$, where  $\alpha >0 , \beta>0$ equal to $\alpha =0.005, \beta = 0.85$.

Figure 1 below displays the dynamic of solution $p(t)$ which satisfies the continuous time SI model (3.4), i.e. with the logistic evolution given in Proposition 2, and  its discrete time Euler approximation at the beginning of an epidemic given in equation(3.2). It is computed with the same values of parameters $\alpha, \beta$ given above. Figure 1 shows that the discrete time approximation (3.2) can be very misleading when it is used for forecasting over a medium or long run.

\medskip
[Insert Figure 1: Evolutions of $p(t)$, SI Model]
\medskip

\nin When $\beta < \alpha$, we get an increasing concave curve that tends to 1. When $\beta > \alpha$, as in Figure 1, we get an exponential convex increase for small $t$, followed by an increasing concave pattern of convergence to 1.

The evolutions of changes of $p(t)$ are shown in Figure 2:

\medskip
[Insert Figure 2: Evolutions of Changes in p(t), SI Model]
\medskip

\nin When $\beta > \alpha$, we get a hump-shaped pattern with the curve decreasing  at a  slower rate after the peak than increasing before the peak.

\nin  We complete the sensitivity analysis of the main features of the SI model by examining the size of  peak (Figure 3) and the time-to-inflection (Figure 4).

\medskip
[Insert Figure 3: Size of Peak, SI Model]

\medskip
[Insert Figure 4: Time to Inflection, SI Model]

\subsection{SIR model}

\nin {\bf 3.2.1 The model}

Let us now consider the SIR model (see Appendix 1) with three states: S for Susceptible; I for Infected, infectious, not immunized; R for Recovered, immunized and no longer infectious. Its transition model representation involves the $3 \times 3$ transition matrix, which is triangular and given by:

row 1, S: $1- \pi(p_2)$; $\pi(p_2)$ ; 0

row 2, I : 0, $p_{22}$; $p_{23}$

row3,R : 0; 0; 1 

\medskip
\nin where $p_{23}$ is strictly positive, and R is the absorbing state.

\medskip
In the limiting case $p_{23}=0$, the $2 \times 2$ North-West subset of the transition matrix corresponds to the SI model discussed in Section 3.1. There are two absorbing states in the SIR model: I and R. State R cannot be reached starting from an initial state S of  susceptible individuals.

\medskip
\nin The marginal probabilities of states I and R satisfy two linearly independent estimating equations:

\begin{eqnarray}
p_2(t) & =  & \pi[p_2(t-1)][1-p_2(t-1)-p_3(t-1)] +p_{22} p_2(t-1), \\
p_3(t) & = & p_{23} p_2(t-1) + p_3(t-1).  \nonumber                                                            
\end{eqnarray}

\nin From the second equation of system (3.5),we get:

\begin{equation}
p_2(t-1)=[p_3(t)-p_3(t-1)]/ p_{23},                                                               
\end{equation}

\nin and by substituting into the first equation, we derive the recursive equation satisfied by $p_3(t)$:

\begin{eqnarray}
p_3(t) &= &p_3(t-1) +  \pi[(p_3(t-1)-p_3(t-2))/p_{23}] [p_{23}-p_3(t-1) +(1-p_{23}) p_3(t-2)]\nonumber \\
& + & p_{22} [p_3(t-1)-p_3(t-2)].   
\end{eqnarray}
\medskip

\nin {\bf Proposition 3} i) The sequence $p_1(t)$ [resp. $p_3(t)$] is decreasing [resp. increasing].

                    ii) The sequence $p_3(t)$ satisfies a nonlinear recursive equation of order 2.

                    iii) The sequence $p_2(t)$ is a linear moving average of order 1 in $p_3(t+1)$.

\medskip

\nin The higher  order of temporal dependence in $p_3(t)$ is due to the interpretation of state I as a transitory state between S and R. Thus, the dynamics of $p_3(t)$ has to account for both the entries into and exits from the state I.

Let us now discuss the behaviour of marginal probabilities p(t) when t tends to infinity. Since $p_3(t)$ is increasing, and it is upper bounded by 1,  its  limit is $p_3(\infty)$, say. From equation (3.6), it follows that the limit of $p_2(t)$ is zero. Then, by taking into account the first equation of (3.5), we get:
 
 \medskip
\nin {\bf Lemma 1:}  When $t \rightarrow \infty$,

i) $p_2(t)\rightarrow 0$.

ii) If $\pi(0)$ is different from 0, $p_3(t) \rightarrow 1$.

iii) If $\pi(0)=0, p_3(t)$ might tend to a limiting value $p_3(\infty) < 1$.

\medskip
Determining the conditions for such a convergence to $p_3(\infty)$, which is strictly less than 1 and computing this limiting value, which is interpreted as the level of {\bf collective immunity} are common topics in the epidemiological literature. Due to the time discretisation bias (see Remark 1), estimation errors on a long run parameter $p_3(\infty)$ are large in an early phase of epidemic. Hence, the estimated ratio of collective immunity and even its existence may be unreliable.

\medskip
\nin {\bf 3.2.2 Homogeneous Markov}

As mentioned Section in 3.1, it is interesting to consider the homogeneous Markov chain, obtained when function $\pi$ is constant (no contagion). Then the evolution of $p(t)$ is driven by a linear recursive equation of order 1: $p(t)= \Pi' p(t-1)$, where $\Pi$ is a triangular matrix with eigenvalues: $1-\pi, p_{22}, 1$. The following proposition is obtained:

\medskip
\nin {\bf Proposition 4}: For a constant $\pi$, we have: $p(t)= A (1-\pi)^t  + B p_{22}^t+ C$,
where $A, B, C$ are 3-dimensional vectors.

\medskip
\nin The effects of entries into and exits from state I induce the two driving exponential functions.
 
\medskip
\nin When function $\pi$ is not constant, the decreasing sequence $p_1(t)$ and  increasing sequence $p_3(t)$ take values between their analogues computed from a homogeneous Markov chain with $\pi=\pi(0)$ and $\pi=\pi(1)$, respectively.

\medskip
\nin {\bf 3.2.3 Local expansion}

At the beginning of an epidemic the probability of being infected $p_2(t)$ is close to 0. Then, the estimating equations can be 
replaced by second-order discrete or continuous deterministic systems of Riccati type. Some of these Riccati systems have closed-form solutions that involve transcendental functions [see Miller (2012), Harko et al. (2014)]. However these analytical solutions have complicated expressions. Alternatively, they can be derived by simulation methods, which are easy to perform in the SIR model.

\medskip

\setcounter{equation}{0}\def\theequation{4.\arabic{equation}}

\nin {\bf 3.2.4 Sensitivity Analysis}

Let us assume a linear function $\pi(p_2) = a + b p_2$, where $a>0, b>0, a+b <1$. Then the SIR model involves 3 independent parameters, and is expected to provide more flexibility than the SI model, due to the additional parameter $p_{23}$. Below, we perform a sensitivity analysis similar to that in Section 3.1.4 and focused on series $p_2(t)$. Parameters $ a, b, p_{23}$ are set equal to $a=0.005, b=0.85, p_{23}=0.5$.

\medskip
[Insert Figure 5: Evolution of $p_2(t)$, SIR Model]

\medskip
[Insert Figure 6 : Evolution of Change in $p_2(t)$, SIR Model]

\medskip
The timing of a peak is determined by parameter $a$ as shown below. We hold parameter $b=0.85$ constant and change the values of parameter $a$ in Figure 7.
\medskip

[Insert Figure 7 : Timing of Peak, $a$ varying, SIR Model]

\medskip
\nin Next, parameter $a=0.005$ is held constant and the values of parameter $b$ are allowed to vary. The size of peak is determined by parameter $b$ as shown in Figure 8 below.

\medskip
[Insert Figure 8: Size of Peak, $b$ varying, SIR model]

\section{Statistical Inference}

This section present the methods of inference for the discrete time transition model.
Let us now consider a parametric transition matrix $\Pi [ p(t-1); \theta]$, with parameter vector $\theta$, and assume that the empirical frequencies $f(t),t=1,..,T$ are observed. In addition, we assume that the parametric model is well specified.

\subsection{Distribution of frequencies}

Under Assumptions A1, A2, these frequencies converge at rate $1/\sqrt{N}$ to their theoretical counterparts and are asymptotically normal. Thus we can write:

\begin{equation}
f(t)=p(t) + u(t),
\end{equation}

\nin where the errors are Gaussian with mean zero and the variance-covariance matrix at lag $h$ given below [Gourieroux, Jasiak (2020)]:

\begin{equation}
Cov[u(t),u(t-h)]=(1/N) \{ \Pi(t-1,h) diag[p(t-h)] - p(t) p(t-h)'\},   
\end{equation}

\nin where: $\Pi(t-1,h)=\Pi[p(t-1)]...\Pi[ p(t-h)]$.

\subsection{(Pseudo) State space representation}

System (4.1) resembles a measurement equation in a state space system with the  measurement variable $f(t)$, measurement error $u(t)$, and the following system of transition equations for the state variable $p(t)$,:

\begin{equation}
p(t)= \Pi[p(t-1); \theta]' p(t-1).  
\end{equation}

However, the system of equations (4.3) and (4.1)  does not fully satisfy the definition of a state space representation because the measurement errors $u(t)$ are 
serially correlated, as shown in equation (4.2).

This difficulty is easily circumvented by assuming a pseudo Gaussian distribution for the errors $u_t'$, and disregarding the autocorrelation. Their variance-covariance matrix at time $t$ can be assumed equal to an identity matrix $Id$ (Ordinary Least Squares approach) or an unknown constant matrix $\Omega$ (Weighted Least Squares approach), or even the true expression of $V(u_t)$ can be considered.  Upon this change of autocovariance structure, a (pseudo) state space representation is obtained.

The pseudo state space representation can be estimated by the Gaussian quasi-maximum likelihood (QML). The quasi-maximum likelihood approach has also an interpretation in terms of estimating equations and asymptotic least squares [see, Berkson (1944), Godambe, Thompson (1974), McRae (1977), Kalbfleisch et al. (1983), Hardin, Hilbe (2003), Miller, Judge (2015)]. As the asymptotic theory is established for $T$ fixed and $N \rightarrow \infty$, the QML and weighted least squares 
methods provide consistent estimators of parameter vector $\theta$, which are not fully efficient as the true structure of autocovariances of errrors $u_t$ has not been taken into account [see, Gourieroux, Jasiak (2020), Appendix 2].

 In practice, the QML estimates \footnote{As the QML approach does not account for the structure of the variance-covariance matrix of the $u(t)$'s it can be improved by replacing a "moment" estimator by a GMM estimator.} of $\theta$ can be computed numerically from an extended, or unscented Kalman Filter \footnote{see, e.g. Song, Grizzle (1995), Krener (2003) for the Extended Kalman Filter.} applied to the (pseudo) state space model.

\medskip
 The estimation approach outlined above remains valid when some frequencies $f_j(t)$  are missing (see Gourieroux, Jasiak(2020) for an application to inference on latent counts of infected and undetected (asymptomatic) individuals).

 The extended Kalman filter provides information on the uncertainty of estimates and predictions. This uncertainty has to be taken into account,  especially at the early ascending phase of an epidemic, when the number of observations is small, the quality of data is rather poor and the parameter of interest, such as the peak, is distant in time from the period of observations. Then, the confidence and prediction intervals are rather wide and the statistician has to interpret the results with caution [see e.g. Viboud et al. (2016), Chowell et al. (2019) for studies at early stages].

 The extended Kalman filter is suitable as an updating algorithm of estimates and forecasts [Nihan, Davis (1987)]. This is especially important at the beginning of the epidemic, when each newly arrived observation is very informative.

\section{Models with Time Dependent Parameters}

\subsection{The modelling}

The simple epidemiological models can be easily extended to allow for time dependent parameters, obtained by replacing $\theta$ by $a(t), b$, say, to distinguish the time dependent parameter vector from the constant parameters. Then, their transition model representation involves the transition matrix $\Pi [p(t-1),a(t),b]$. The time dependent propagation parameters can, for instance, capture the time varying  implementation of  social distancing measures and the compliance with these measures [see e.g. DiDomenico et al. (2020), Alvares et al. (2020)]. Such an extended model can be written also in the (pseudo) state space representation and estimated by using the methods given in the previous Section. The (pseudo) state space representations depend on the assumptions on the evolution of parameter vector $a(t)$. At least three types of modelling approaches can be considered.

\medskip
i) {\bf Dynamics of $a(t)$ left unspecified}

\nin The state space representation comprises the measurement equation:
$$f(t)=p(t) + u(t), $$

\nin and the transition equations:

$$p(t)= \Pi[ p(t-1), a(t), b]' p(t-1).$$

\nin The state variables are the marginal probabilities $p(t)$ and parameters $a(t)$ while $b$ is the vector of constant parameter. They can be jointly estimated and the state variables filtered by the extended Kalman filter, under an identification condition. In particular, the order condition: $(J-1)(T-1) \geq  T \, dim \, a + dim \, b$ has to be satisfied.

\medskip

ii) {\bf Stochastic evolution of $a(t)$}

\nin An alternative model has the same measurement equation and extends the previous state space representation system as it includes additional transitions such as:

$$a(t)= \phi a(t-1) + v(t),$$

\nin where the errors $v(t)$ are Gaussian noises independent of the measurement errors $u(t)$ and $|\phi|<1$ for stationarity.

\nin Under this representation, the state variables are the marginal probabilities $p(t)$ and parameters $a(t)$, and $b$ is the constant parameter vector. The extended Kalman Filter can be used to jointly estimate $b$ and filter the components of $a(t)$.

\medskip
iii) {\bf Exogenous information on $a(t)$}

If the indicators $x(t)$ of social distancing are available, such as counts of travellers [see, Hurtacsu et al. (2020)] and the daily numbers of fines for disobeying the social distancing rules, the model can be extended to include $x(t)$. Then, the model is similar to the representation above with the autoregressive dynamic replaced by an equation such as:

$$a(t)=C x(t) + v(t)$$

\nin where $C$ is a row vector of constant coefficients to be estimated and $v(t)$ are Gaussian noises independent of the measurement errors $u(t)$.
\medskip

\medskip

\nin {\bf 5.2 Logistic model with stochastic contagion parameter}

Let us now examine the limiting case of time independent stochastic parameters, which is 
a very special case of stochastic dynamics and consider the logistic model of Proposition 2. The stochastic parameters are introduced to account for heterogeneity of infection patterns. For ease of exposition, we assume a discrete heterogeneity distribution with weights $q_k, k=1,...,K$ on values $(\alpha_k, \beta_k), \;k=1,...,K$ [see, Gourieroux et al.(1996), Yan, Chowell (2019)]. Then, equation (3.4) is written conditional on $\alpha, \beta$ and defines the evolution of $p_{2,k}(t)$ for values $\alpha_k, \beta_k$. Next, these evolutions need to be re-integrated  with respect to $\alpha, \beta$, in order to find the marginal probability of being infected $p_2(t)$ as follows:

$$ p_2(t) = \sum_{k=1}^K q_k p_{2,k}(t) = \sum_{k=1}^K  \{ q_k [\alpha_k [\exp (\alpha_k + \beta_k)t] - \alpha_k]\, /\, [\alpha_k \exp [(\alpha_k + \beta_k) t ] +\beta_k] \}. $$

\nin We get a convex combination of logistic functions \footnote{used as a basis of functions in neural networks.}. This additive specification implies that $p_2(t)$ cannot follow a quadratic differential equation such as (3.4). There is a double heterogeneity i) in coefficients $\alpha$, which means that there exist multiple initial exogenous clusters of infection of different sizes, ii) in coefficients $\beta$, which means that the speeds of propagation of each cluster are different. This model is not a special case of (SI)$^K$
[see Appendix 1] as there is no contagion between the sub-populations $k$ and only contagion within is allowed.

The presence of heterogeneity in a logistic model generates the following effects:

i) several peaks can appear in the changes in $p_2(t)$ over time. This is the wave effect [Witham (1974), Gourieroux et al. (1996)], due to different propagation parameters of each wave.

ii) the persistence of $p_2(t)$ increases due to the additive representation. This is a well-known long memory effect revealed in Granger, Joyeux (1980) for linear autoregressive models that also exist in nonlinear logistic models [see Sattenspiel (1990)].

Below, a three-wave pattern is illustrated in Figure 9 for the following
parameter values: $\alpha = 0.015, 0.0005, 0.0001$, $\beta = 0.95, 0.85, 0.75$

\medskip

[Insert Figure 9: Three-Wave Infection Pattern]

\medskip
We observe the three waves with the highest peak due to the first wave. We also observe a persistence effect as the decline following the first peak is slower than the declines of each wave separately.

\section{Concluding Remarks}

Contrary to the major part of literature on epidemiological models, which considers deterministic continuous time models of counts of individuals in various compartments (states), we consider stochastic models in discrete time for variables representing individual histories [see also Allen (1994), Das et al.(2011) for discrete time approach]. The proposed discrete time transition model has the following advantages:

i) It eliminates the lack of consistency between time discretized continuous time models with respect to the time unit. In the continuous time setup, it also eliminates the assumption of differentiability of aggregate counts, which are discrete variables. This is especially important in the early phase of an epidemic when some of these counts are small.

 ii) The stochastic model allows us to combine aggregate count variables and individual medical histories of patients under medical care.
 
 iii) The stochastic component allows us for deriving not only the point, but also interval forecast.  This is important at the beginning of an epidemic when the number of observations is small and the results are less reliable.

iv) The estimation of the transition model can be performed by applying an extended Kalman filter to its (pseudo) state space representation.

\newpage

{\bf {\Large REFERENCES}} \\

\nin Admani,J., Hattaf, K.,and N., Yousfi (2018); Dynamical Behaviour of a Stochastic SIRS Epidemic Model,Journal of Mathematical and Computational Science,8,421-431. \\

\nin Allen, L. (1994): Some Discrete-Time SI, SIR  and SIS Epidemic Models, Mathematical Biosciences, 124, 83-105. \\

\nin Alvarez, F., Argente, D. and F. Lippi (2020): A Simple Planning Problem for
COVID-19 Lockdown, LUISS and Einaudi Institute for Economics and Finance DP. \\

\nin Atkeson, A. (2020): What Will Be the Economic Impact of Covid-19 in the US? Rough Estimates of Disease Scenarios, NBER Working Paper 26867. \\

\nin Berkson, J. (1944): Application of the Logistic Function to Bio-Assay, JASA, 339- 357.\\

\nin Brandenburg,H.(2019):Quadratic Growth During the 2019 Novel Coronavirus Epidemic,arXiv.\\

\nin Brauer, F. and C. Castillo-Chavez (2001): Mathematical Models in Population Biology and Epidemiology,Springer,New York. \\

\nin Chowell, G., Tariq, A., and J.M. Hyman (2019):A Novel Sub-Epidemic Modeling Framework for Short Term Forecasting Epidemic Waves, BMC Med.,17,1-19. \\

\nin Das, P., Mukherjee, D., and A., Sarkar (2011): Study of an SI Epidemic Model with Nonlinear Incidence Rate: Discrete and Stochastic Version, Applied Mathematics and Computation,218,2509-2515.\\

\nin DiDomenico, L., Pullano, G., Coletti, P., Hens, N. and V. Colizza (2020):Expected Impact of School Closure and Telework to Mitigate COVID-19 Epidemic in France, www.epicx-lab.com/covid-19.html]. \\

\nin El Koufi, A., Adnani, J., Bennar, A. and N. Yousfi (2019): Analysis of a Stochastic SIR Model with Vaccination and
Nonlinear Incidence Rate, International Journal of Differential Equations,Vol 2019, ID 9275051. \\

\nin Feng, Z., Huang, W., and C, Castillo-Chavez (2005): Global Behaviour of a Multi Group SIS Epidemic Model with Age Structure, Journal of Differential Equations, 218,292-314.\\

\nin Godambe, V. and M. Thompson (1974): Estimating Equations in the Presence of a Nuisance Parameter, Annals of Statistics, 3, 568-576. \\

\nin Gourieroux, C., and J.,Jasiak (2006): Multivariate Jacobi Process with Smoothed Transition, Journal of Econometrics,131,475-505. \\

\nin Gourieroux, C. and J. Jasiak (2007): Econometrics of Individual Risks: Credit, Insurance and Marketing, Princeton University Press. \\

\nin Gourieroux, C. and J.,Jasiak (2020): Time Varying Markov Processes with Partially Observed Aggregate Data: An Application to Coronavirus, Arxiv 2005.04500.  \\

\nin Gourieroux, C. and I. Peaucelle (1996): Diffusion and Wave Effects, Annales d'Economie et de Statistique, 44, 191-217.\\

\nin Granger, C. and A. Joyeux (1980): "An Introducction to Long Memory Time Series Models and Fractional Differencing", Journal of Time Series Analysis, 1 , 15-29.

\nin Hardin, J. and J. Hilbe (2003): "Generalized Estimating Equations", Chapman Hall. \\

\nin Harko, T., Lobo, F. and M. Mak (2014): Exact Analytical Solutions of the Susceptible-Infected-Recovered Epidemic Model and of the SIR  Model with Equal Death and Birth Rates,Applied Mathematics and Computation,236,184-194. \\

\nin Harvey, A., and P. Kattuman (2020): "Time Series Models Based on Growth Curves with Applications to Forecasting Coronavirus", Covid Economics, 24, 126-157. \\

\nin Hethcote, H. (2000): The Mathematics of Infectious Diseases,SIAM Review, 42, 599-653. \\

\nin Hortescu, A., Liu, J. and T., Schwieg (2020): Estimating the Fraction of Unreported Infections in Epidemics with a Known Epicenter:An Application to COVID-19,DP University Chicago. \\

\nin Jiang, D., Yu, J., Ji, C. and N., Shi (2011): Asymptotic Behaviour of Global Positive Solution to a Stochastic SIR Model,Mathematical and Computer Modelling,54,221-232. \\

\nin Kalbfleisch, J., Lawless, J., and W. Vollmer(1983): Estimation in Markov Models from Aggregate data,Biometrics,39,907-919. \\

\nin Kermack, W., and A. McKendrick (1927): A Contribution to the Mathematical Theory of Epidemics,Proceedings of the Royal Statistical Society, A, 115,700-721. \\

\nin Korolev, I. (2020): Estimating the Parameters of the SEIRD Model
for COVID-19 Using the Deaths Data, Department of Economics, Binghamton University DP. \\

\nin Krener, A. (2003): The Convergence of the EKF, Directions in Mathematical Systems: Theory and Optimization, 173-182, Springer. \\

\nin Kuhi, D., Kebraeb, E., Lopez, S. and J., France (2003): An Evaluation of Different Growth Functions for Describing the Profile of Live Weight with Age in Meat and Egg Strains of Chicken,Poultry Science,82,1536-1543.\\

\nin McFadden, D. (1984): Econometric Analysis of Qualitative Response Models,in Handbook of Econometrics,Vol 2,1395-1457, Elsevier. \\

\nin McRae, E. (1977): Estimation of Time Varying Markov Processes with Aggregate Data,Econometrica, 45,183-198. \\

\nin Meng, X., and L., Chen (2008): The Dynamics of a New SIR Epidemic Model Concerning Pulse Vaccination Strategy,Applied Mathematics and Computation,197,582-597.\\

\nin Miller, J. (2012): A Note on the Derivation of Epidemic Final Sizes,Bulletin of Mathematical Biology, 74, Section 4.1. \\

\nin Miller,J. and G. Judge (2015): "Information Recovery in a Dynamic Statistical Markov Model", Econometrics, Vol 3/2, 187-198. \\ 

\nin Nihan, N. and G., Davis (1987): Recursive Estimation of Origin-Destination Matrices From Input/Output Counts, Transpn. Research,218,149-163.\\

\nin Richards, F. (1959): A Flexible Growth Function for Empirical Use",J.Exp.Bot.,10,290-301. \\

\nin Sattenspeil, L. (1990): Modeling the Spread and Persistence of Infectious Disease in Human Populations", Yearbook of Physical Antropology.\\

\nin Song, Y. and J. Grizzle (1995): The Extended Kalman Filter as a Local Asymptotic Observer, Estimation and Control, 5, 59-78. \\

\nin Toda, A. (2020): Susceptible-Infected-Recovered (SIR): Dynamics
of COVID-19 and Economic Impact, ArXiv:2003.11221v2 \\

\nin Tchuenche,J., Nwagwo,A and R;, Levins (2007): Global Behaviour of an SIR Epidemic Model with Time Delay, Mathematical Methods in Applied Sciences,30,733-749.\\

\nin Verity et al. (2020): Generalized Estimates of the Severity of Coronavirus Disease 2019: A Model Based Analysis, online. \\

\nin Viboud C., Simonsen, L., Chowell, G. (2016): A Generalized Growth Model to Characterize the Early Ascending Phase of Infectious Disease Outbreaks,Epidemics,15,27-37. \\

\nin Vynnicky,.E., and White,R. (eds)(2010): An Introduction to Infectious Disease Modelling, Oxford Univ. Press. \\
 
\nin Witham, G. (1974): Linear and Nonlinear Waves, Wiley. \\ 
 
\nin Wu, K., Darcet, D., Wang, Q. and D., Sornette (2020): Generalized Logistic Growth Modelling of the COVID-19 Outbreak in 29 Provinces in China and in the Rest of the World,DP ETH Zurich .\\ 
 
\nin Wu, K., Zheng, J., and J.,Chen (2020):Utilize State Transition Matrix Model to Predict the Novel Coronavirus Infection Peak and Patient Distribution, ArXiv. \\

\nin Zhang,I.,and Z., Ma (2003): Global Dynamics of a SEIR Epidemic model with Saturating Contact Rate,Mathematical Biosciences,185,15-32.\\

\nin Yan, P., and G. Chowell (2019): Quantitative Methods for Investigating Infectious Disease Outbreaks, Springer.

\newpage

\bc{\bf Appendix A.1}\ec 
 
\bc {\bf Structural Epidemiological Models}
\ec

\nin This Appendix presents a typology of basic  epidemiological models that can serve as building blocks of more complex specifications. The difference between the basic models are with respect to:

\medskip

i) the number of states ( compartments) and their interpretations as S= susceptible, E=exposed, I =infected, R =recovered, D=deceased.

ii) The number and types of virus propagation sources

iii) The location of zeros in the transition matrix (i.e. the causal structure)

iv) The structure of time dependent transition probabilities.

\nin We provide below the transition probabilities along with the state interpretations. The time dependent transition probabilities are denoted by $\pi$ and are functions of (lagged) marginal probabilities $p(t)$.

\nin The models described below are the following:

2-state: SI model, SIS model

3-state:SIR model

4-state : SIRD model, SEIR model,  SIR model, (SI)$^2$ model, S,IU ,ID, R model.

\nin The interpretations of  states and the form of transition matrices are given below.
\medskip

\nin 2-state SI model

\nin S= susceptible,I= infected,being immunized and staying infectious for the S people

\nin row 1,S: $\pi_{11}(p_2), \pi_{12}(p_2)$

\nin row 2, I : 0,1

\nin One absorbing state,one source of infection.

\medskip
\nin 2-state SIS model

\nin S:susceptible, I infected, can recover, but without being immunized.

\nin row 1,S: $\pi_{11}(p_2)$, $\pi_{12}(p_2)$

\nin row 2,I: $p_{21}, p_{22}$, with $p_{21}>0$. 

One source of infection, no absorbing state;a non degenerate stationary solution
can exist [see e.g. Allen (1994), Feng et al (2005) for the use of SIS model].

\medskip
\nin 3-state SIR model

\nin S=susceptible, I= infected, infectious, not immunized, R=recovered, no longer infectious,immunized.

\nin row1,S: $\pi_{11}(p_2), \pi_{12}(p_2), \pi_{13}(p_2)$

\nin row 2, I: 0, $p_{22}, p_{23}$

\nin row 3, R: 0,0,1

\nin One absorbing state,one source of infection [see e.g. Tchenchue et al.(2003), Meng,Chen(2008),Jiang et al. (2011), Toda (2020)].

\medskip
\nin 4-state SIRD model

\nin S=susceptible, I=infected, not immunized,infectious, R=recovered, no longer infectious, immunized, D=deceased.

\medskip
\nin row 1, S: $\pi_{11}(p_2), \pi_{12}(p_2), \pi_{13}(p_2), p_{14}$

\nin row 2, I : 0, $p_{22}, p_{23}, p_{24}$

\nin row 3, R: 0,0, $p_{33}, p_{34}$

\nin row 4, D: 0,0,0,1

\nin One absorbing state, one propagation source.

\medskip
\nin 4-state (SI)$^2$

\nin The population is divided into 2 sub-populations, as region 1 \& region 2, male \& female, young \& old. It is easily extended to any number of regions.

\medskip
\nin $S_j$ =susceptible of type $j$, $I_j$= infected, immunized, infectious of type $j$.

\nin row 1, $S_1$: $\pi_{11}(p_3,p_4), 0, \pi_{13}(p_3,p_4),0$

\nin row 2, $S_2$: 0, $\pi_{22}(p_3,p_4), 0, \pi_{24}(p_3,p_4)$

\nin row 3, $I_1$: 0,0,1,0

\nin row 4,$I_2$: 0,0,0,1

\nin Two absorbing states,two propagation sources  [see e.g. Feng et al. (2005)].

\medskip

\nin 4-state SEIR

\nin S=susceptible
E= exposed, but not yet infectious (there is a latency period), not immunized,
I = infected and infectious, not immunized,
R=recovered,no longer infectious,immunized.

\medskip
\nin row 1,S: $\pi_{11}(p_3), \pi_{12}(p_3), 0, 0$ 

\nin row 2,E: 0, $p_{22}, p_{23}, 0$

\nin row 3,I: 0, 0, $p_{33}, p_{34}$

\nin row 4,R: 0,0,0,1

\nin One absorbing state,one propagation source [see e.g. Zhang,Ma (2003)].

\medskip

\nin 4-state: S IU ID R

\nin S=susceptible

\nin IU:infected,infectious, not immunized,undetected

\nin ID: infected,infectious, not immunized,detected

\nin R:recovered,no longer infectious, immunized

\medskip
\nin row 1, S : $\pi_{11}(p_2,p_3), \pi_{12}(p_2,p_3), \pi_{13}(p_2,p_3), p_{14}$

\nin row 2, IU: 0, $p_{22}, p_{23}, p_{24}$

\nin row 3, ID: 0, 0, $p_{33}, p_{34}$

\nin row 4, R: 0, 0, 0, 1

\nin One absorbing state, two propagation sources [see e.g. Gourieroux, Jasiak (2020)].
\medskip

These structural models can be extended by considering other states, such as the birth
\footnote{In transition models the state birth is usually introduced to balance the deaths and to provide stationary evolutions of the processes [see e.g. Harko et al. (2014)]. This ad-hoc introduction of births is not relevant at the beginning of the epidemic when the  interest is in determining the nonstationary dynamic at the beginning of the epidemic, rather than in the long run equilibrium. Second, there is no increased count of births (known 9 months earlier) in order to offset the 
increasing number of deaths due to coronavirus.} to offset the future number of deaths due to coronavirus, the types of medical treatment of patients in  hospitals, the severity (asymptomatic (mild), symptomatic, high), or the detection (contact tracing, influenza like illness surveillance, tests, etc) [see, e.g. Verity et al. (2020)]. The models can also be extended by combining the basic models as building blocks to construct a 5-state model such as the SEIRD [see e.g. Korolev(2020)], or S IU ID R D, a 6-state model such as the 
S$IU^2$, $ID^2$ , R, or $(SIR)^2$. The structure of the transition matrix can also be modified to account for the possibility that a fraction of recovered individuals is not entirely immunized and can be infected twice.

\medskip
\newpage

\bc{\bf Appendix A.2}\ec 
 
\bc {\bf Rational Recursive Equations}
\ec

This Appendix shows the exact solution and the exact time discretization of the probability to be infected in a continuous time SI model. We provide below the results for a general one-dimensional Riccati equation in continuous time written on a series $x(t)$. Then, the results can be applied to the special case $x(t)= p_2(t)$.
\medskip

\nin i) {\bf The differential equation}

\nin This differential equation is:

$$ dx(t)/dt= - \lambda (x(t)-a) (x(t)-b) /(a-b), $$

\nin where $\lambda$ is strictly positive.

\medskip

\nin ii) {\bf The solution}

\nin Since: 

$$ (a-b)/[(x(t)-a)(x(t)-b)]= 1/(x(t)-a) -1/(x(t)-b), $$ 

\nin we deduce:

$$ dx(t)[1/(x(t)-a) -1/(x(t)-b)]=- \lambda dt,$$

\nin and by integration:

$$ |x(t)-a|/|x(t)-b|= exp(- \lambda t) |x(0)-a|/|x(0)-b|.$$

The form of this relation implies that the trajectory $x(t)$ and the starting value $x(0)$ satisfy always the same relationship with respect to a and b, that is $x(t)$ is in the interval (a,b) (resp. below, above), if $x(0)$ is in this interval (resp. below, above). Therefore, we can disregard the absolute values to get:

$$ (x(t)-a)/(x(t)-b) =exp(- \lambda t) (x(0)-a)/(x(0)-b),$$ 

\nin for any nonnegative $t$.

This implies a logistic expression for $x(t)$:

$$ x(t)= [a-b k exp(-\lambda t)]/[1- k exp(-\lambda t)],$$

\nin where: $k= (x(0)-a)/(x(0)-b)$.

\medskip
\nin iii) {\bf The exact time discretized recursive model}

We deduce that the exact time discretized counterpart corresponds to a rational transform of $x(t-1)$. More precisely, we get:

$$x(t)= \{ a[x(t-1)-b]-b[x(t-1)-a] exp(-\lambda )\}/\{[x(t-1)-b]-[x(t-1)-a] exp(-\lambda)\}.$$

\nin This rational recursive equation, which is the exact time discetization, differs from the crude Euler discretization.

\medskip
\nin iv) {\bf Special case}

The results above can be computed for equation (3.4) with:
$ \lambda= \beta >0$, $a=- \alpha/\beta, b=1$. Without an exogenous source of infection: $\alpha=0$ ,the interval $(a,b)$ is the interval (0,1), $x(t)=p_2(t)$ is decreasing and tends to 0 when t tends to infinity, for any starting value $p_2(0)$.

\newpage

\begin{center}
\begin{figure}[h]
\centering
\includegraphics[width=12cm,angle = 0]{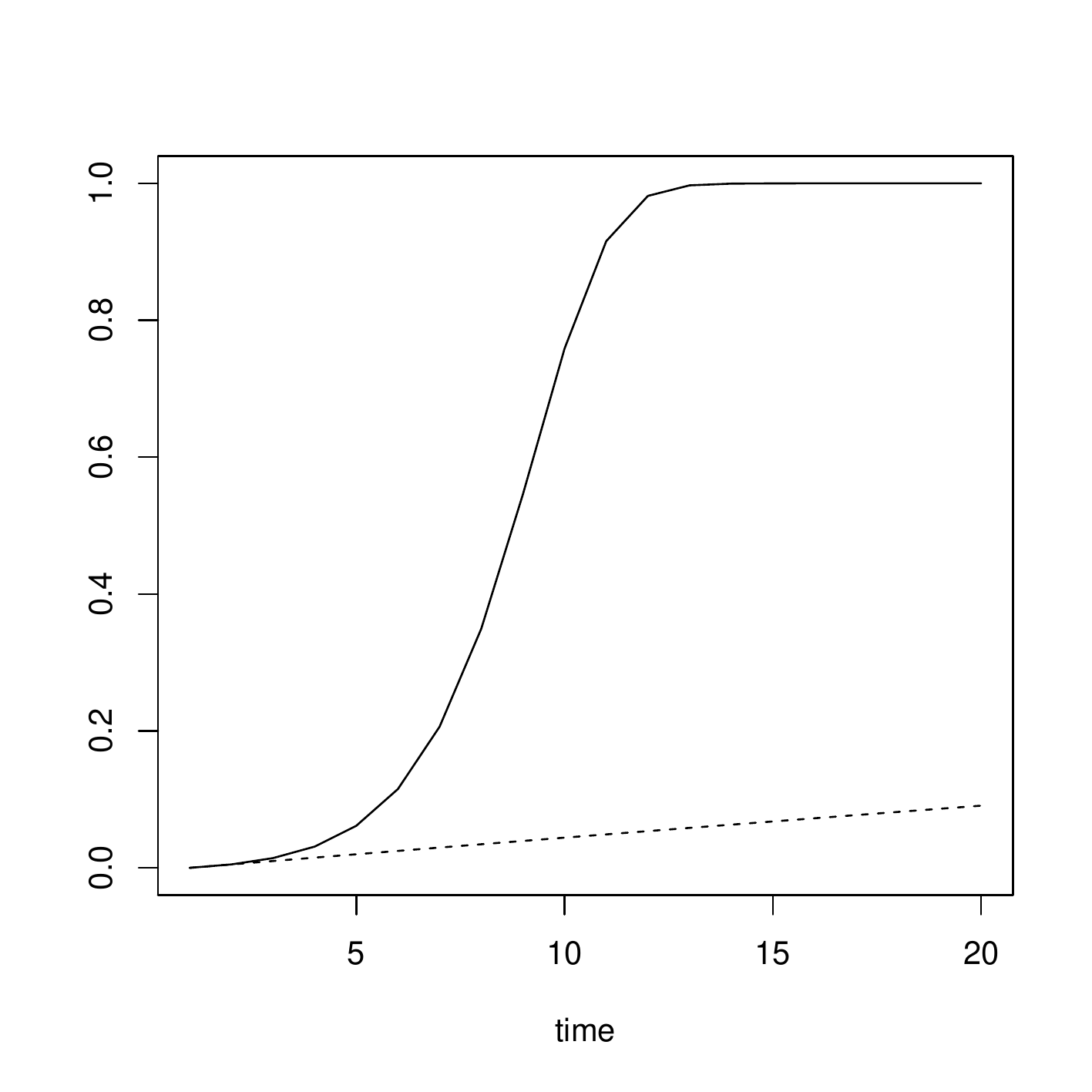}
\caption{Evolution of $p(t)$, SI Model}
\end{figure}
\end{center}

\newpage

\begin{center}
\begin{figure}[h]
\centering
\includegraphics[width=12cm,angle = 0]{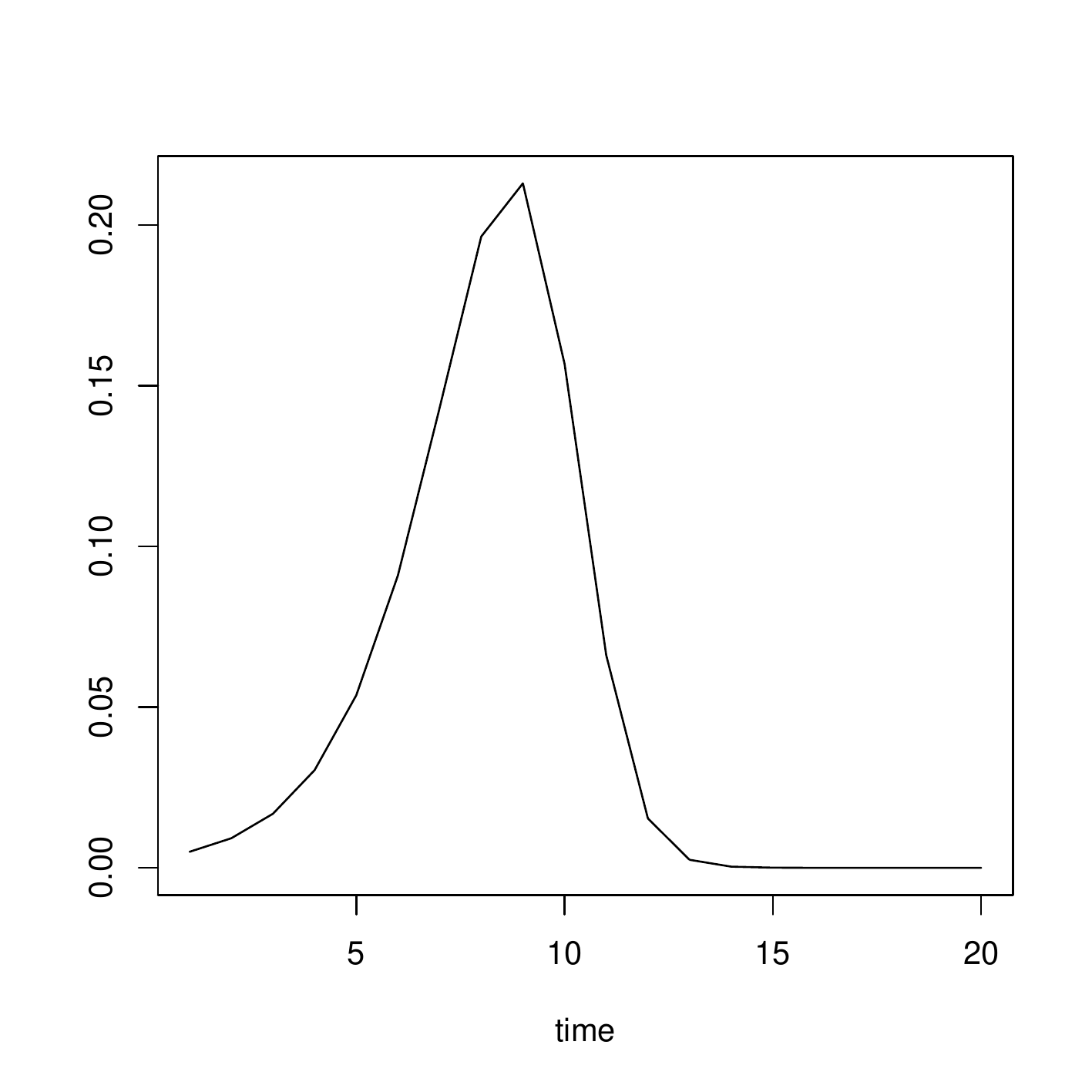}
\caption{Evolution of Changes in $p(t)$, SI Model}
\end{figure}
\end{center}

\newpage

\begin{center}
\begin{figure}[h]
\centering
\includegraphics[width=12cm,angle = 0]{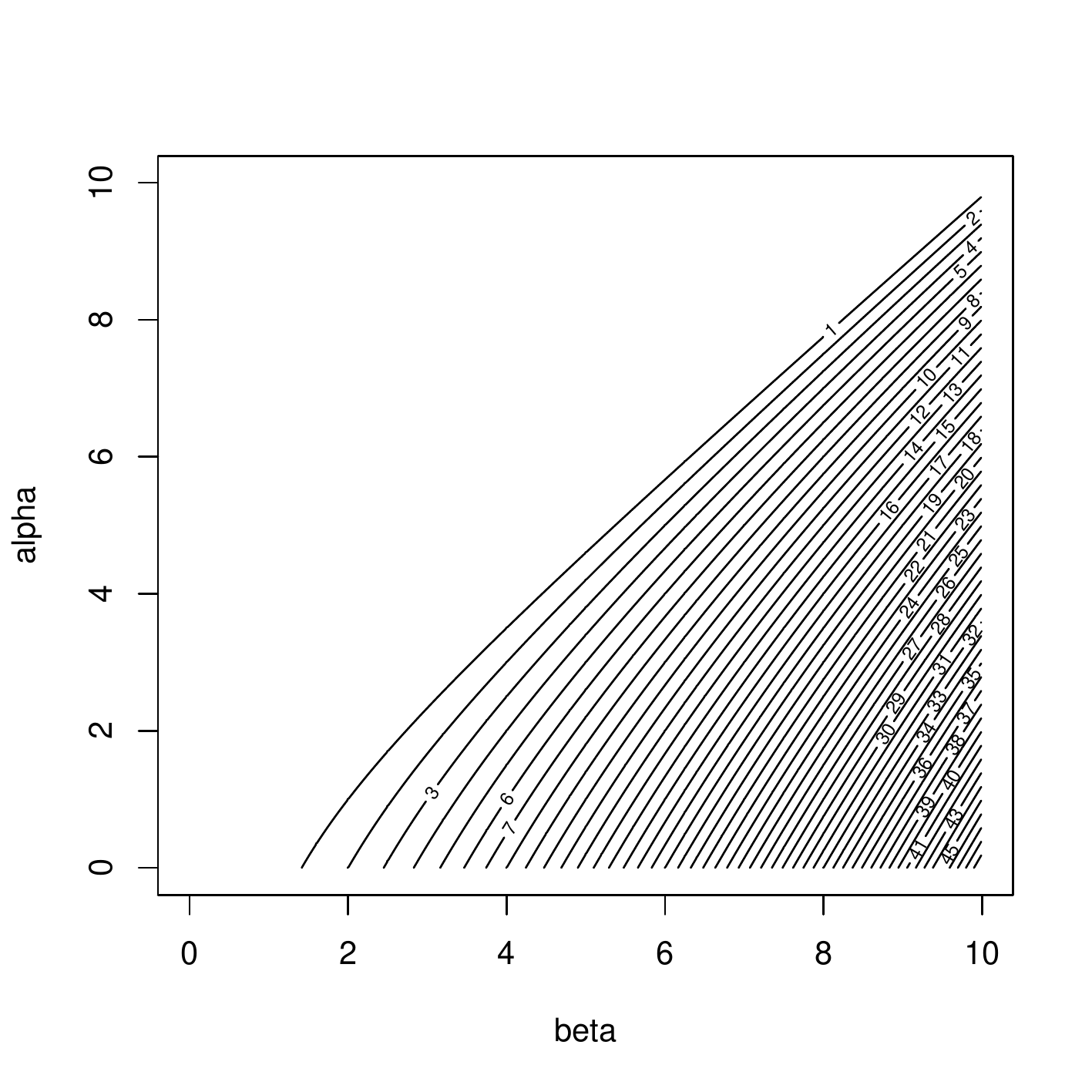}
\caption{Size of Peak, SI Model}
\end{figure}
\end{center}

\newpage

\begin{center}
\begin{figure}[h]
\centering
\includegraphics[width=12cm,angle = 0]{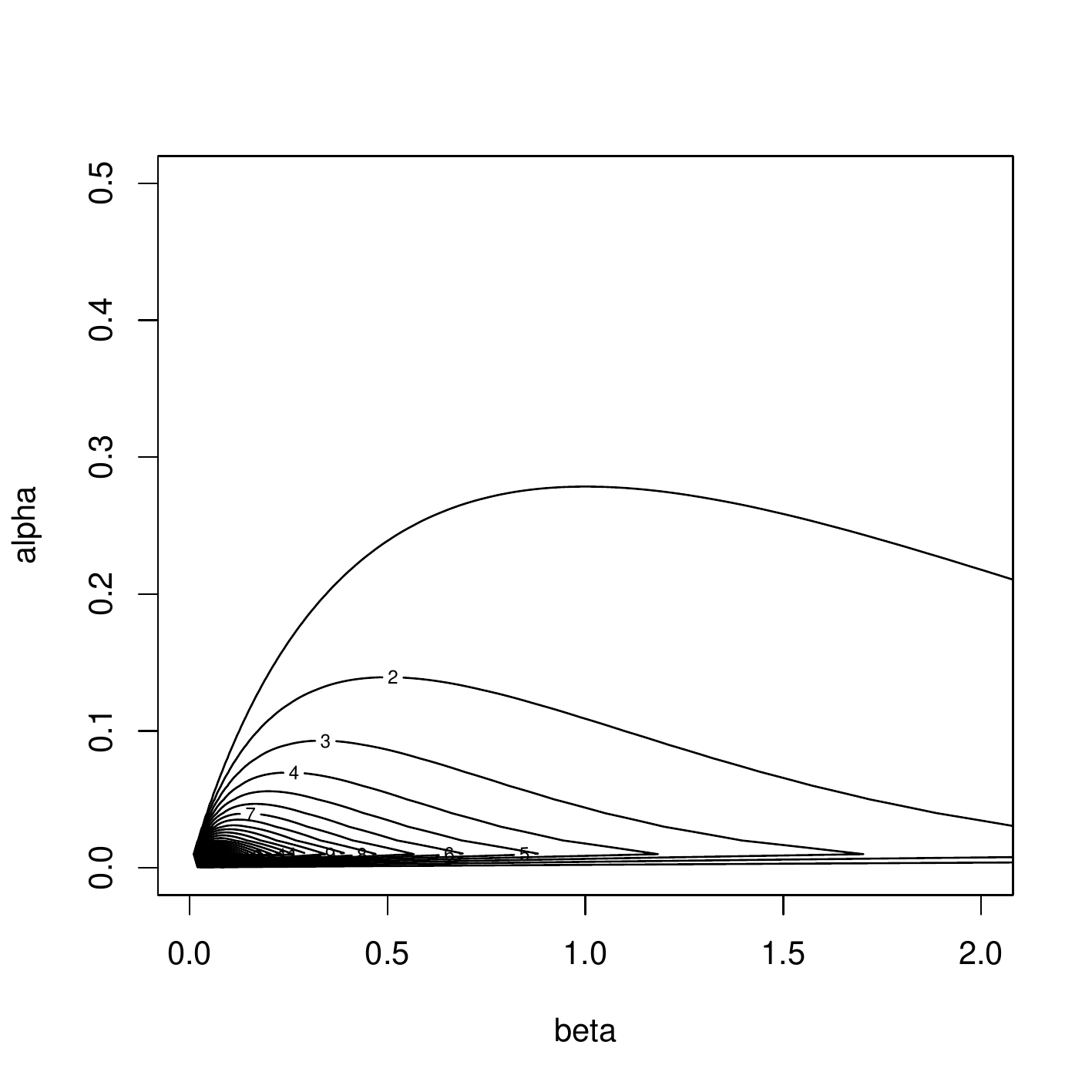}
\caption{Time to Inflection, SI Model}
\end{figure}
\end{center}

\newpage

\begin{center}
\begin{figure}[h]
\centering
\includegraphics[width=12cm,angle = 0]{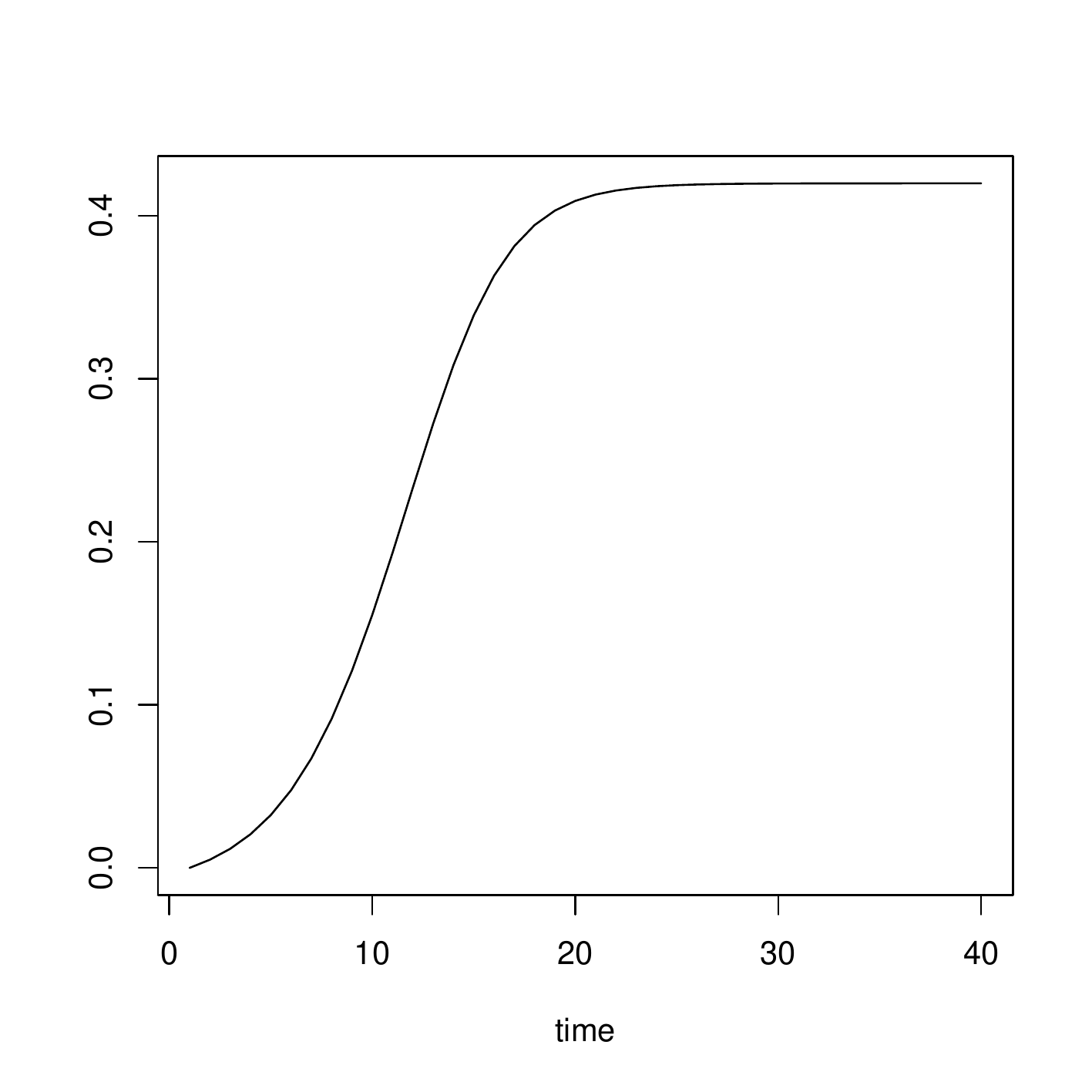}
\caption{Evolution of $p_2(t)$, SIR Model}
\end{figure}
\end{center}

\newpage

\begin{center}
\begin{figure}[h]
\centering
\includegraphics[width=12cm,angle = 0]{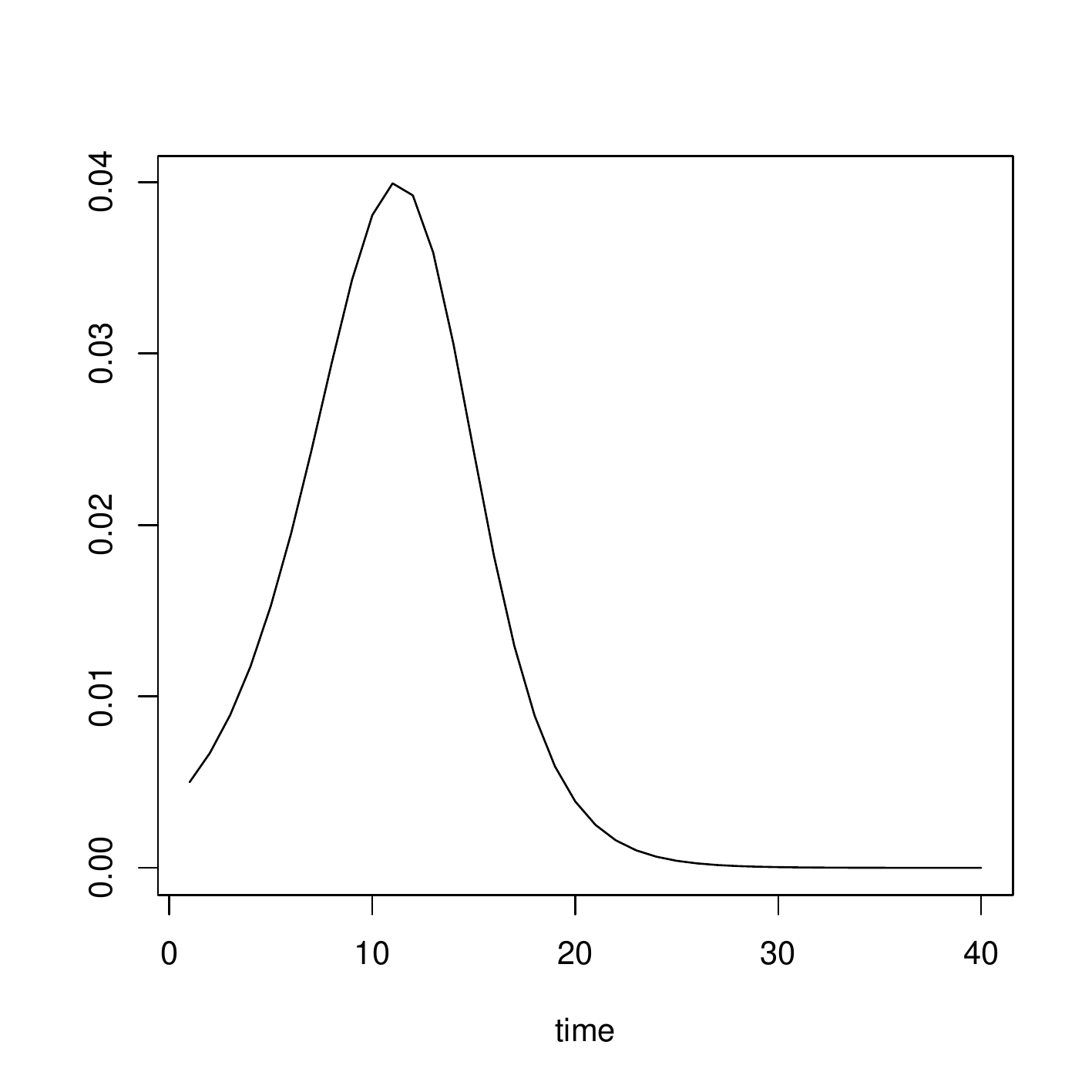}
\caption{Evolutions of Changes in $p_2(t)$, SIR Model}
\end{figure}
\end{center}

\newpage

\begin{center}
\begin{figure}[h]
\centering
\includegraphics[width=12cm,angle = 0]{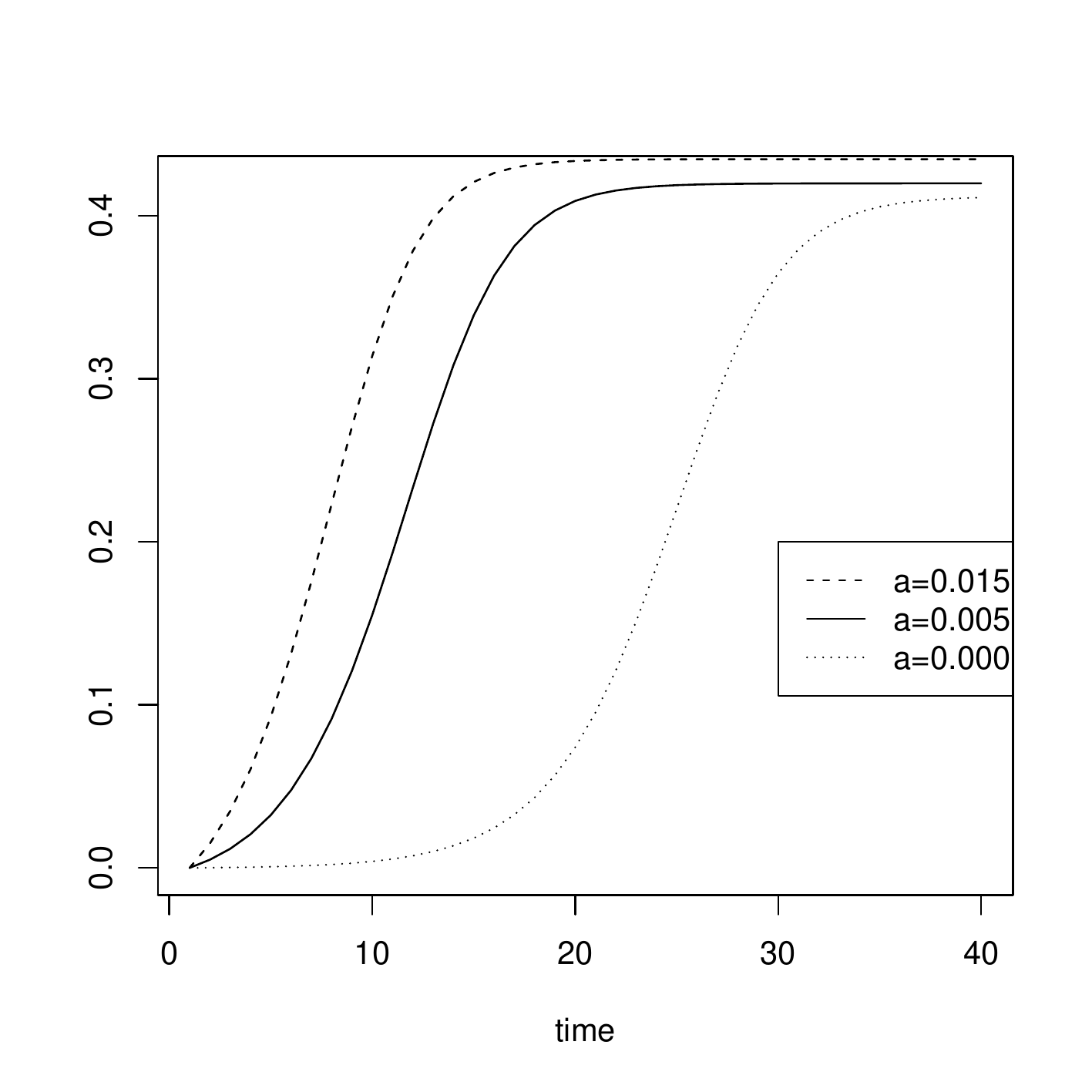}
\caption{Timing of Peak, $a$ varying, SIR Model}
\end{figure}
\end{center}

\newpage

\begin{center}
\begin{figure}[h]
\centering
\includegraphics[width=12cm,angle = 0]{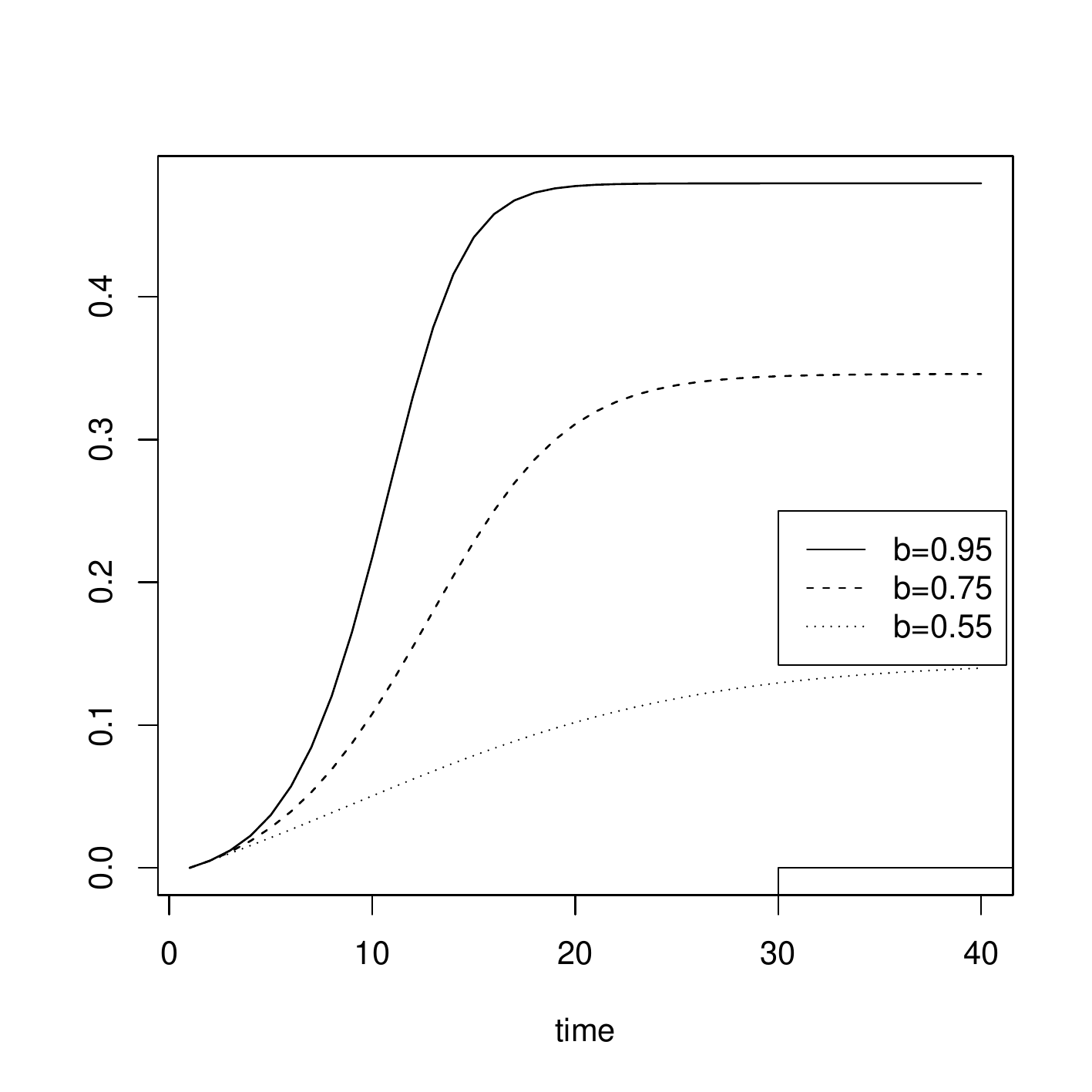}
\caption{Size of Peak, $b$ varying, SIR Model}
\end{figure}
\end{center}

\newpage

\begin{center}
\begin{figure}[h]
\centering
\includegraphics[width=12cm,angle = 0]{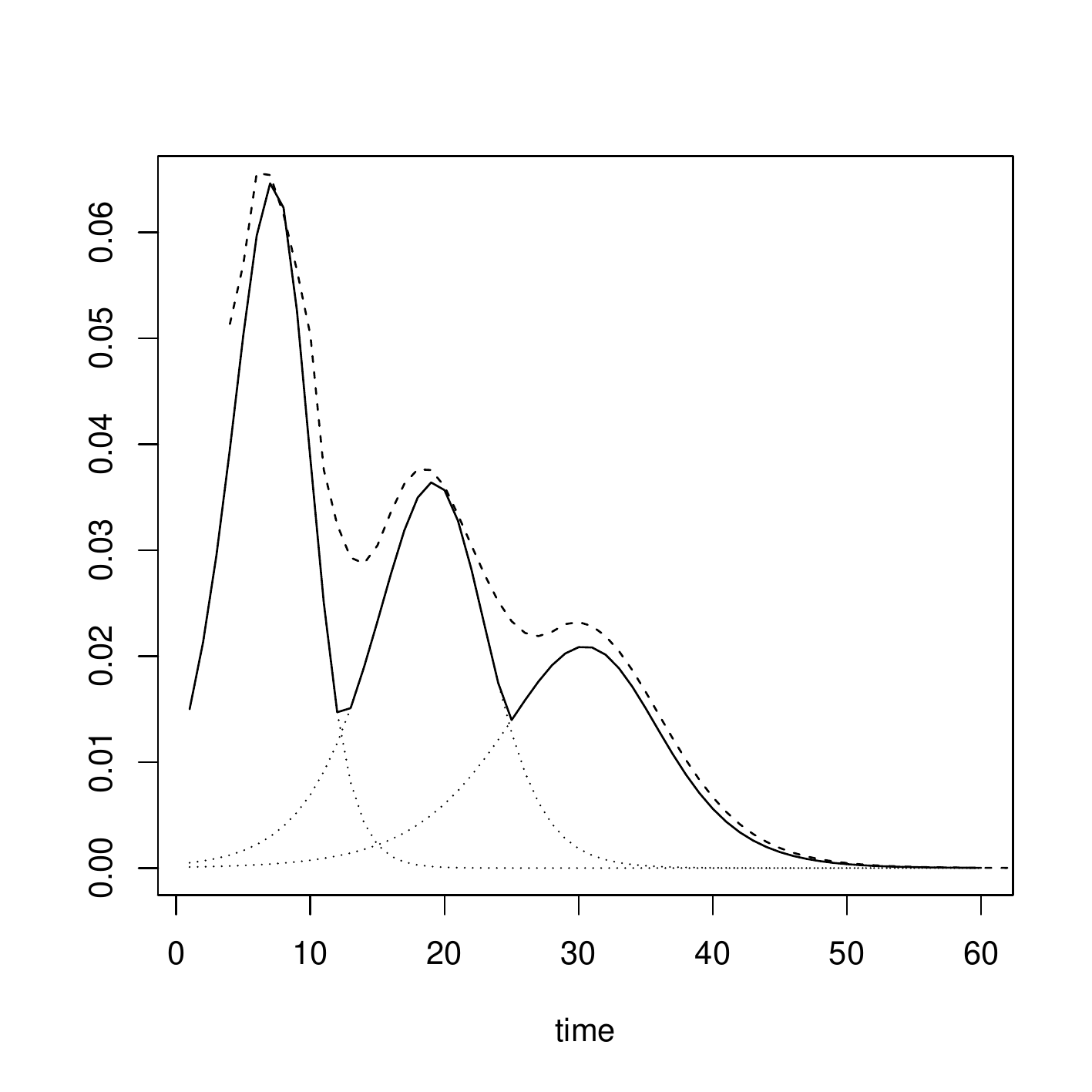}
\caption{Three-Wave Infection Pattern}
\end{figure}
\end{center}
\end{document}